\documentclass[conference]{IEEEtran}
\IEEEoverridecommandlockouts
\usepackage{cite}
\usepackage{amsmath,amssymb,amsfonts}
\usepackage{algorithmic}
\usepackage{graphicx}
\usepackage{textcomp}
\usepackage{xcolor}
\usepackage{svg}
\usepackage{adjustbox}
\makeatletter
\newcommand{\linebreakand}{%
  \end{@IEEEauthorhalign}
  \hfill\mbox{}\par
  \mbox{}\hfill\begin{@IEEEauthorhalign}
}
\makeatother

\begin{document}
\bstctlcite{IEEEexample:BSTcontrol}

\title{Protecting Voice-Controlled Devices against LASER Injection Attacks
}

\author{\IEEEauthorblockN{Hashim Ali, Dhimant Khuttan, Rafi Ud Daula Refat, and Hafiz Malik}
\IEEEauthorblockA{\textit{Department of Electrical Engineering}\textit{University of Michigan-Dearborn, Dearborn, MI USA}\\
% \\
\{alhashim, dkhuttan, rerafi, hafiz\}@umich.edu}
}

\maketitle
% \textcolor{blue}{text}
\begin{abstract}
Voice-Controllable Devices (VCDs) have seen an increasing trend towards their adoption due to the small form factor of the MEMS microphones and their easy integration into modern gadgets. Recent studies have revealed that MEMS microphones are vulnerable to audio-modulated laser injection attacks. This paper aims to develop countermeasures to detect and prevent laser injection attacks on MEMS microphones. A time-frequency decomposition based on discrete wavelet transform (DWT) is employed to decompose microphone output audio signal into n + 1 frequency subbands to capture photo-acoustic related artifacts. Higher-order statistical features consisting of the first four moments of subband audio signals, e.g., variance, skew, and kurtosis are used to distinguish between acoustic and photo-acoustic responses. An SVM classifier is used to learn the underlying model that differentiates between an acoustic- and laser-induced (photo-acoustic) response in the MEMS microphone. The proposed framework is evaluated on a data set of 190 audios, consisting of 19 speakers. The experimental results indicate that the proposed framework is able to correctly classify $98\%$ of the acoustic- and laser-induced audio in a random data partition setting and $100\%$ of the audio in speaker-independent and text-independent data partition settings.
\end{abstract}


\section{Introduction}
MEMS microphones are becoming a de facto standard for voice-activated devices due to their small form factor, easy integration in analog-to-digital converter (ADC) chips, and reduced interference and SNR problems. As a consequence, markets for MEMS microphones have expanded rapidly from around 433 million units in 2009 (\$2 billion USD) to 4.65 billion units in 2016 (\$5 billion USD) \cite{bogue2013recent, zawawi2020review}. Today, the top 3 products that employ MEMS microphones are mobile handsets, media tablets, and wearable electronics, of which 1.35 billion mobile handsets were sold around the world in 2020. This market penetration shows the importance of MEMS microphones for years to come.

This wide-scale adoption of MEMS microphones has turned a large array of applications into Voice-Controllable (VC) systems. Today, people can use Apple Siri and Google Home to initiate calls, find the location of a parked car, open/close garage doors, control lighting at home, etc. \cite{abidi2018development, sen2015design}; whereas Amazon Alexa \cite{amazonAmazonAlexa} has also allowed users to buy things online using voice commands. Furthermore, financial institutions are also keen to integrate financial services with these Voice-Controllable devices \cite{zou2021design,zhang2020artificial}.

Despite its numerous benefits, the MEMS microphone is adding a new attack surface to Voice-Controllable systems. Recently, a research group from the University of Michigan has successfully exploited the MEMS microphone-induced attack surface by injecting audio into a Voice-Controllable device using LASER \cite{sugawara2020light}. This laser injection attack method modulates the recorded audio of the target speaker onto a laser and directs the laser to a MEMS microphone of the VC system. The MEMS microphone demodulates the audio signal and feeds it to the underlying audio processing pipeline. The LASER injection attack has added a new tool to an attacker's already rich tool set, which now enables him to execute audio commands from a distance of up to 100 m \cite{sugawara2020light}.

% [trim={left bottom right top},clip]
\begin{figure*}[!h]
\centering
\adjustbox{trim=0 2.3cm 0 2.3cm}{%
 \includegraphics[width=1\linewidth]{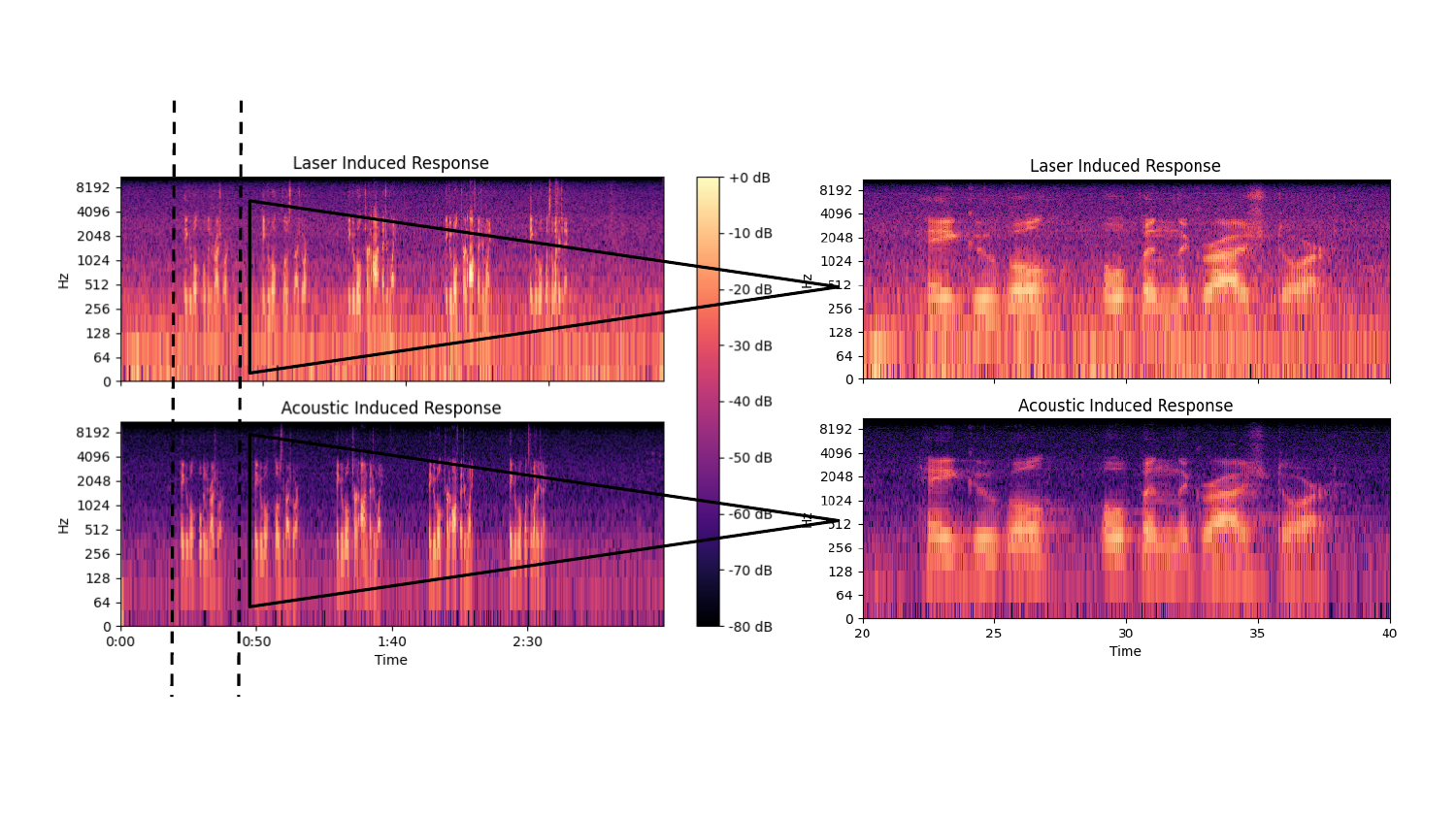}
}

\caption{Spectral Analysis}
\label{fig:spec_analysis}
\end{figure*}

This paper aims to develop countermeasures to detect and prevent laser injection attacks. To the best of our knowledge, this is the first attempt that investigates the laser (photo-acoustic) response in the microphone and proposes a framework to detect the signal-level characteristics of the laser-induced response. The proposed method relies on the hypothesis that the acoustic-induced activity response in the microphone is different from the laser-induced response, which can be leveraged to detect a laser attack on the VADs. Our initial investigation suggests that laser-induced response exhibits unique artifacts in the response signal that can be observed in the spectral analysis as shown in figure \ref{fig:spec_analysis}. It can be observed from figure \ref{fig:spec_analysis} that the laser-induced response exhibits noise in the low-to-mid-frequency region which is absent in the acoustic-induced response. In addition, acoustic-induced response exhibits sound reflection properties such as echo and reverberation, which is absent in the laser-induced response. We leveraged these artifacts in section \ref{proposed} to develop a countermeasure for detecting laser-injection attacks to MEMS microphone-based Voice-Controllable devices.\\

The various ways a MEMS microphone-based Voice-Controllable system can be attacked is discussed in Section \ref{threat}, and their proposed countermeasures in the literature in Section \ref{related_work}. After that, a machine learning-based countermeasure that can detect a laser-induced response in the microphone is described in Sections \ref{proposed}. In the end, the proposed countermeasure is evaluated on a dataset of acoustic-induced and laser-induced audio recordings, the results of which are discussed in Section \ref{exp_res}.

\section{Threat Model} 
\label{threat}

\begin{figure}[h]
\centering
\includegraphics[width=1\linewidth]{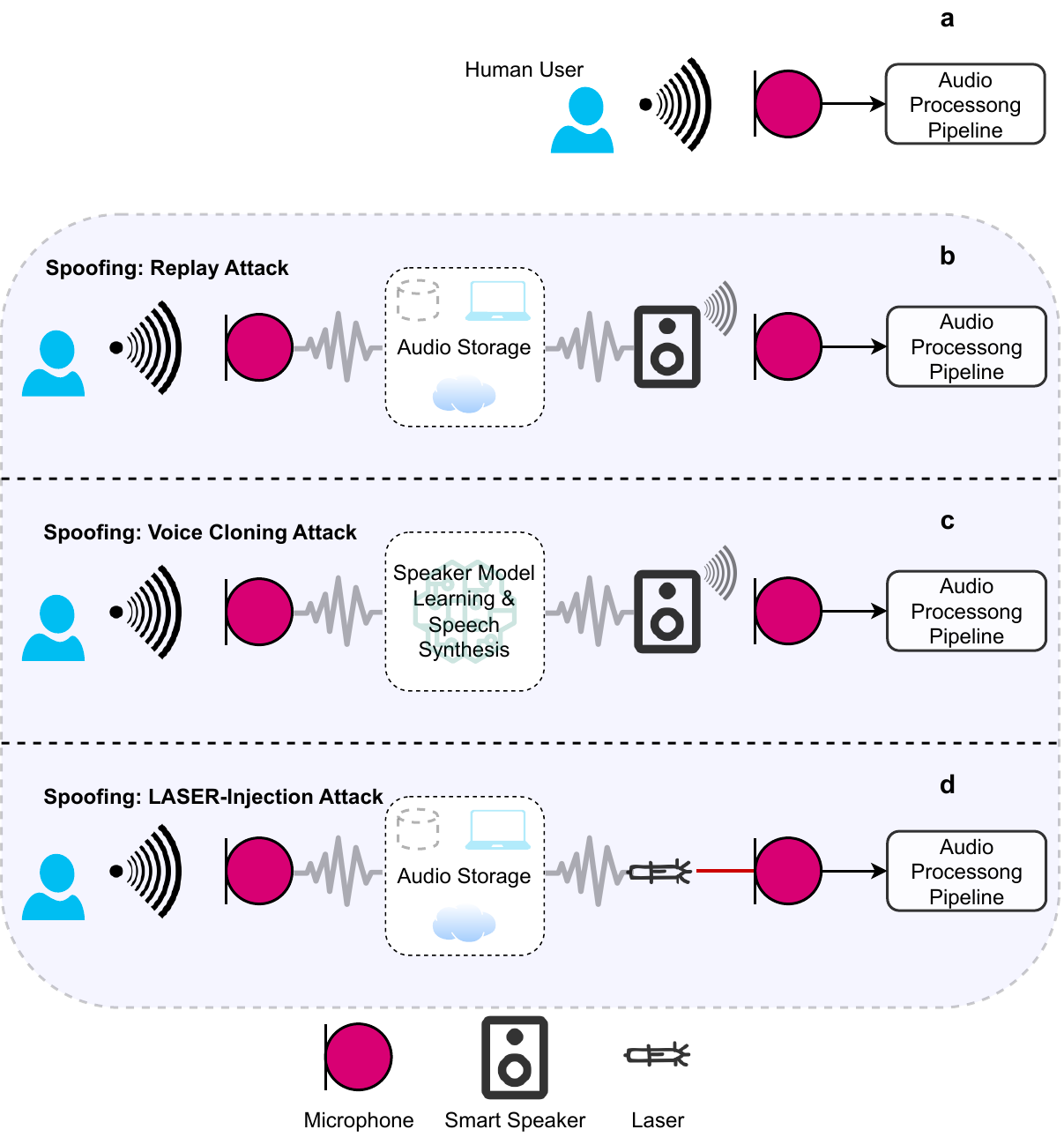}
\caption{Different Threat Models of a Voice Controllable System}
\label{fig:types_attacks}
\end{figure}

Voice-Controllable (VC) devices are vulnerable to various spoofing attacks, including replay, cloning, and LASER injection attacks. When an attacker uses the recorded voice of a target speaker and replays it in front of a VC device for illicit purposes, it is termed a replay attack (see Fig. \ref{fig:types_attacks} -b). On the other hand, when an attacker uses machine learning algorithms to generate synthetic audios of a target speaker and inject them directly into a VC device - bypassing the microphone, it is termed a voice cloning attack (please see Fig \ref{fig:types_attacks}-c). The third type of attack, as successfully demonstrated by Sugawara et al. \cite{sugawara2020light}, is called a LASER-injection attack. This attack is similar to a replay attack in the sense that a target speaker's recorded audio is modulated onto the laser, and that laser is injected into the microphone. This attack can be launched from up to a distance of 100m (please see Fig \ref{fig:types_attacks}-d). According to AsvSpoof \cite{kamble2020advances, nautsch2021asvspoof}, spoofing attacks on MEMS microphone-based Voice Controllable devices can be divided into two main categories: (i) attacks based on physical access (PA) and (ii) attacks based on logical access (LA). Physical access (PA)-based attack is an attack in which an attacker needs physical access to the microphone to launch the attacks. On the other hand, logical access (LA) involves attacks that are injected directly into the VC device bypassing the microphone. As replay and LASER-injection attacks require physical access to the microphone therefore these attacks can be categorized as PA-based attacks. The proposed threat modeling of VC devices is shown in Figure \ref{fig:types_attacks}.

% [trim={left bottom right top},clip]

\section{Background and Related Work} 
\label{related_work}

To the best of our knowledge, no countermeasure exists for detecting LASER injection attacks. Nonetheless, this section provides a brief description of countermeasures proposed in the literature for protecting Voice-Controllable systems from other types of Voice Spoofing attacks \cite{kamble2020advances}. Existing countermeasures to voice spoofing attacks can be broadly classified into two categories: 

\begin{itemize}
    \item Classical Machine Learning Approaches
    \item Representation Learning Approaches
\end{itemize}

Classical machine learning-based countermeasures for audio spoof detection typically consist of two parts. The first part deals with hand-crafted feature extraction and the second part consists of a model that determines the authenticity of the audio signal \cite{balamurali2019toward}. In the context of feature extraction, researchers have proposed various acoustic features to counter voice spoofing attacks. Cepstral coefficient features including constant-Q transform (CQT), Log-CQT, constant-Q cepstral coefficient (CQCC), extended CQCC (eCQCC), inverted CQCC (iCQCC), linear frequency cepstral coefficient (LFCC), Mel-frequency cepstral coefficient (MFCC) have been used widely \cite{todisco19_interspeech, das2019long, cai2019dku, yang2019sjtu, adiban2020replay}. The second part of classical machine learning approaches consists of a model that determines the authenticity of the audio signal. These models have evolved over the last 40 years. For example, earlier researchers used systems based on Discrete Vector Quantization \cite{mishra2012vector}. After that, the state of the art moved to solutions based on the Gaussian mixture model (GMM) \cite{reynolds2000speaker}, and more recently moved to i-vector frameworks based on factor analysis \cite{matrouf2007straightforward}. However, Gaussian Mixture Model-based systems are still more common among researchers and industry professionals. For example, Asvspoof Challenge 2019 \cite{nautsch2021asvspoof} recommended two baseline systems for the performance evaluation of the algorithms participating in the challenge. Baseline 1 used Q cepstral coefficients (CQCCs) as features \cite{todisco19_interspeech} and baseline 2 used linear frequency cepstral coefficients (LFCCs) \cite{sahidullah2015comparison} as features. Both baselines employed the Gaussian mixture model as a classifier to protect and prevent the MEMS microphone-based VC devices from voice spoofing attacks.

Representation learning approaches work in the form of feature learning \cite{qian2016deep} or as a pattern classifier \cite{yu2017spoofing}. However, it was observed that the use of deep neural networks followed by a classifier such as GMM or SVM performs better than just using deep neural networks as classifiers \cite{kamble2020advances}. In such approaches, hidden layers perform a feature extraction task, and then a GMM or SVM classifier performs the classification task \cite{chen2015robust}. For instance, RNN features followed by a GMM classifier resulted in 2.5\% EER for all kinds of attacks in AsvSpoof 2015 \cite{qian2016deep,kamble2020advances}. Chen et al. \cite{chen2017resnet} proposed a fusion of GMM, DNN, and Resnet classifiers on MFCC and CQCC features to detect voice replay attacks. This method achieved a 13.3\% EER on the Asvspoof 2017 evaluation dataset.

% \subsection{Background} \label{bg}
Sugawara et al. in \cite{sugawara2020light} presented the laser injection attack that has its basis in photoacoustic effect, which has been studied quite extensively \cite{manohar2016photoacoustics}. The first work in the area of photo-acoustics dates back to 1800 when Alexander Graham Bell invented a device that used a vibrating mirror and a selenium cell to modulate sunlight and convert it to electricity. However, the rise of digital communication technology and the need to have a line of sight between the transmitter and receiver made this technology less attractive. More recently, researchers rediscovered voice-over light transmission and Patrick Tucker reported the development of a device by the US military that ionizes molecules in the air to generate sound. Infrared laser-based sound generation was proposed in \cite{sullenberger2019photoacoustic} which can deliver sound over a range of 2.5m.

\begin{figure}[h]
\centering
\includegraphics[width=1\linewidth]{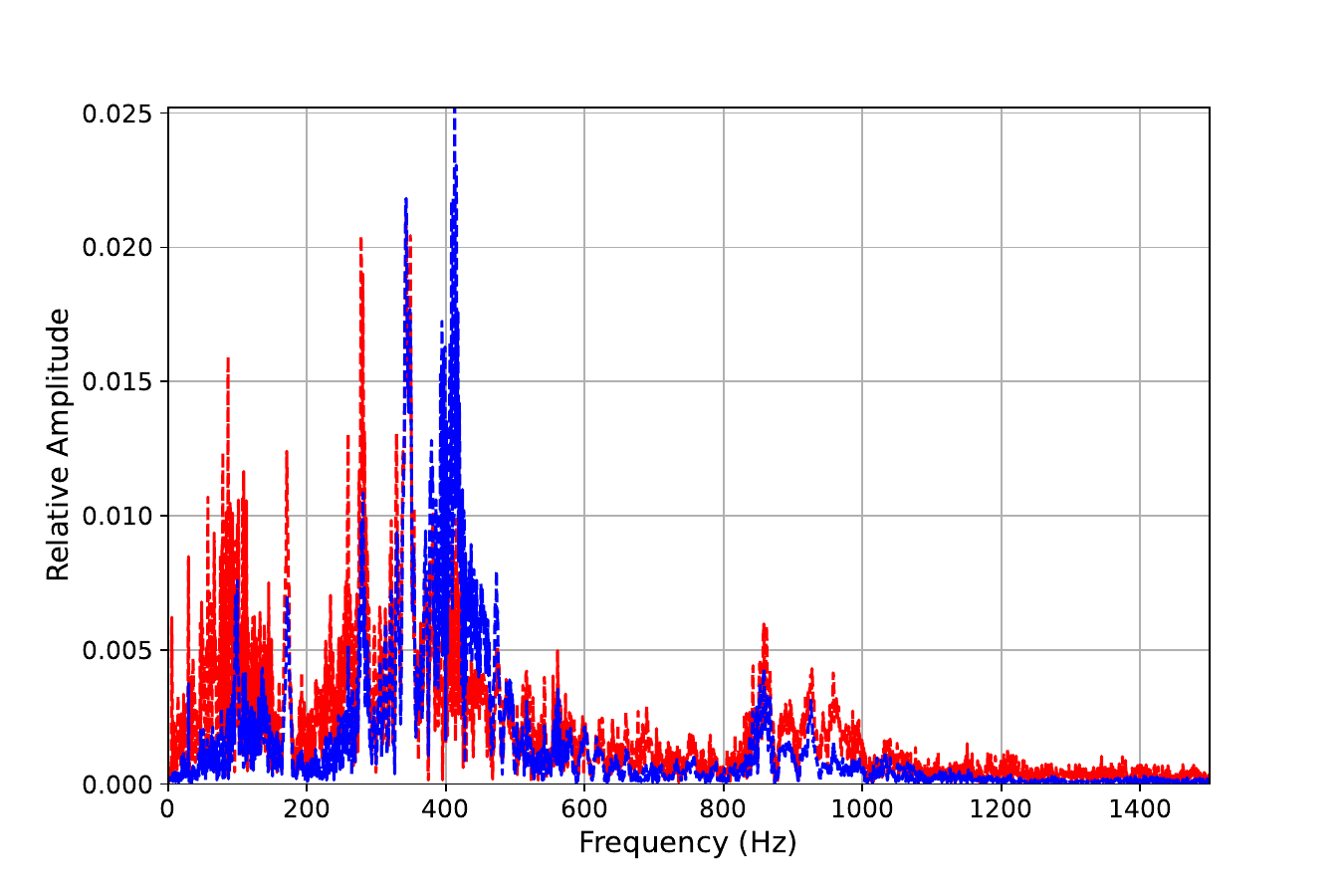}
\caption{``Hey Google'' Spectrum. Acoustic-induced Audio (blue) vs. Laser-induced Audio (red)}
\label{fig:Hey_Google_spectrum}
\end{figure}

The initial analysis in Sugawara et al. \cite{sugawara2020light} suggested that the laser injection attack might exploit the photoacoustic and photoelectric phenomenon. Therefore, the authors of \cite{sugawara2020light} sought to characterize the laser injection attack on the Mems microphone in a new study \cite{cyr2021lasers}. Similarly, a 2021 master's thesis at Linkoping University investigated laser and ultrasonic injected signals in microphones \cite{djerv2021investigation}. The investigation of Laser-injection attack on MEMS microphones performed in \cite{cyr2021lasers} and \cite{djerv2021investigation} is in line with our initial hypothesis that the laser (photo-acoustics) induced response in the microphone exhibits low-frequency noise or in other words, low frequencies are dominant in the laser-induced response and higher frequencies are suppressed. To strengthen our confidence in this hypothesis, we injected ``Hey Google'' through a laser into the microphone, and recorded the laser-induced and acoustic-induced responses of the microphone in a frequency vs. amplitude graph as shown in Figure \ref{fig:Hey_Google_spectrum}. This figure shows that the lower frequencies are dominant in the laser-induced microphone response compared to the acoustic-induced microphone response.

\begin{figure*}[!h]
\centering
\includegraphics[width=0.8\linewidth]{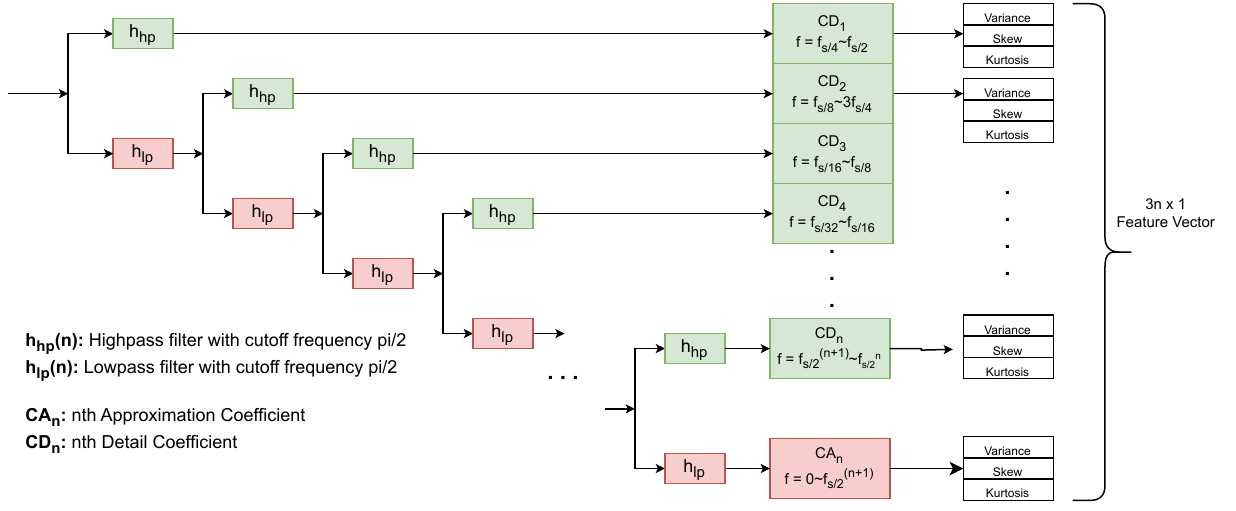}
\caption{Discrete Wavelet Transform (DWT) Based Countermeasure to Laser-Injection Attack}
\label{fig:dwt}
\end{figure*}

\section{Proposed Framework} 
\label{proposed}
As discussed in section \ref{related_work} that low frequencies are dominant in the laser (photo-acoustic) induced response in the microphone and higher frequencies are suppressed. A classical machine learning-based approach can be applied where features are extracted from the acoustics-induced audio and laser-induced audio and then a classification model can be employed to differentiate between two types of audio. To take advantage of this difference between the acoustic-induced response and laser-induced response, a filter bank based on discrete wavelet transform (DWT) is implemented, which splits the incoming audio signal into $n+1$ frequency subbands. At each subband level, the low-frequency sub-band is further decomposed into low and high-frequency sub-bands called Approximation and Detail Coefficients respectively. At level $n$, we have one approximation coefficient array and $n$ detail coefficient arrays (n-level decomposition is shown in figure \ref{fig:dwt}). \textbf{It is important to note here that the DWT-based approach, as described in this section, is not the only approach that can detect laser-injection attacks on MEMS microphone-based VC devices}. Other approaches, as demonstrated in the experimental section (section \ref{exp_res}), may also work. However, no such demonstration exists for these approaches, which is the purpose of this paper.

A distribution is fitted to each sub-band coefficient array of both acoustic-induced and laser-induced audio. The Approximation Coefficient array, $CA_5$ follows a Lognorm distribution for laser-induced audio and a Cauchy distribution for acoustic-induced audio. All detail coefficient arrays, $CD_5 - CD_1$, follow a Cauchy distribution with laser-induced audio exhibiting shorter peaks as compared to acoustic-induced audio. A distribution plot of the Approximation Coefficient, $CA_5$, and the very next Detail Coefficient, $CD_5$, for acoustic-induced audio and laser-induced audio is shown in figure \ref{fig:dwt}.

\begin{figure}[h]
\centering
\includegraphics[width=1\linewidth]{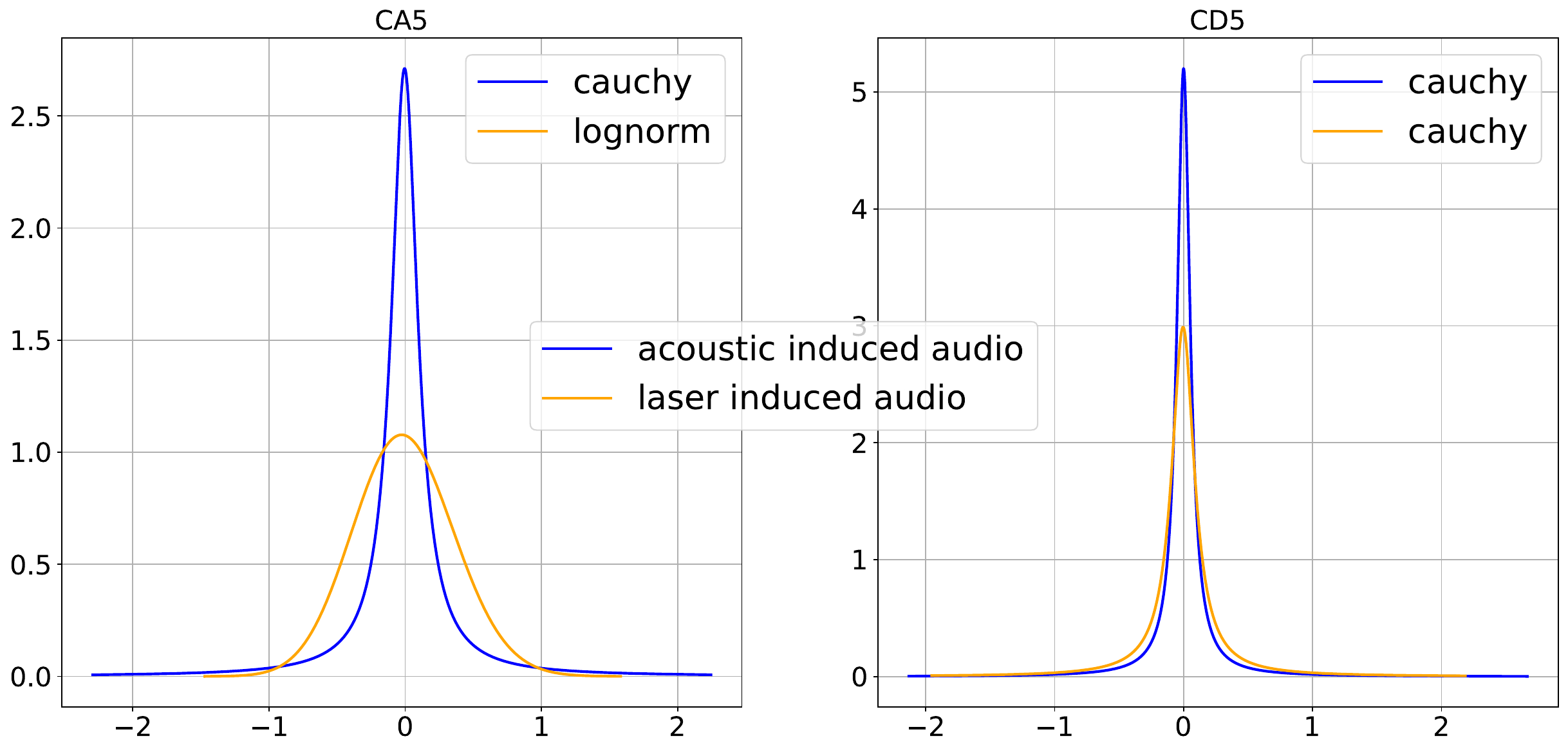}
\caption{\textbf{$CA_5$:} laser-induced audio follows a Lognorm distribution, acoustic-induced audio follows a Cauchy distribution. \textbf{$CD_5$:} Both laser-induced and acoustic-induced audio follow a Cauchy distribution}
\label{fig:distri_coeff}
\end{figure}

Based on the distribution plots from figure \ref{fig:distri_coeff}, laser artifacts can be captured through higher-order statistical features of the approximation and detail coefficient arrays. Therefore, the second, third, and fourth moments (also known as variance $\sigma$, skew $s$, and kurtosis $\kappa$) of these coefficient arrays are computed. These higher-order statistical features, variance $\sigma$, skew $s$, and kurtosis $\kappa$ are concatenated into a feature vector of size $3(n+1)\times1$, where $n$ is the level that the incoming audio signal is decomposed into. Consequently, a machine learning model was trained on this feature vector to learn the underlying structure that differentiates acoustic-induced audio from laser-induced audio. The detail of the proposed method for the decomposition level equal to $n$ is shown in figure \ref{fig:dwt} above. $CD_n ... CD_1$ are the detail coefficient arrays and $CA_n$ is the approximation coefficient array. 

\begin{table*}
\begin{center}
\caption{Comparison of DWT Features with existing baseline features.\\
    \textbf{SD}: Speaker Dependent, \textbf{SI}: Speaker Independent, \textbf{TD}: Text Dependent, \textbf{TI}: Text Independent }
\resizebox{0.8\textwidth}{!}{%
    \begin{tabular}{ p{3cm} p{3cm} p{3cm} p{3cm} }
     \hline
     % \multicolumn{4}{|c|}{} \\
     % \hline
     Method  &  Accuracy (SD + TD)  &  Accuracy (SI + TD)  &  Accuracy (SI + TI)\\
     \hline
     \textbf{DWT + SVM}   &  \textbf{0.98}  &  \textbf{1.0}   &  \textbf{1.0}\\
     CQCC + SVM  &  0.91  &  1.0   &  1.0\\
     \textbf{LFCC + SVM}  &  \textbf{0.98}  &  \textbf{1.0}   &  \textbf{1.0}\\
     MFCC + SVM  &  0.96  &  0.96  &  0.95\\
     \hline
    \end{tabular}%
}
    \label{table:table_comp}

\end{center}
\end{table*}

\section{Experimentation and Results} 
\label{exp_res}
\begin{figure}[h]
     \centering
     \includegraphics[width=1\linewidth]{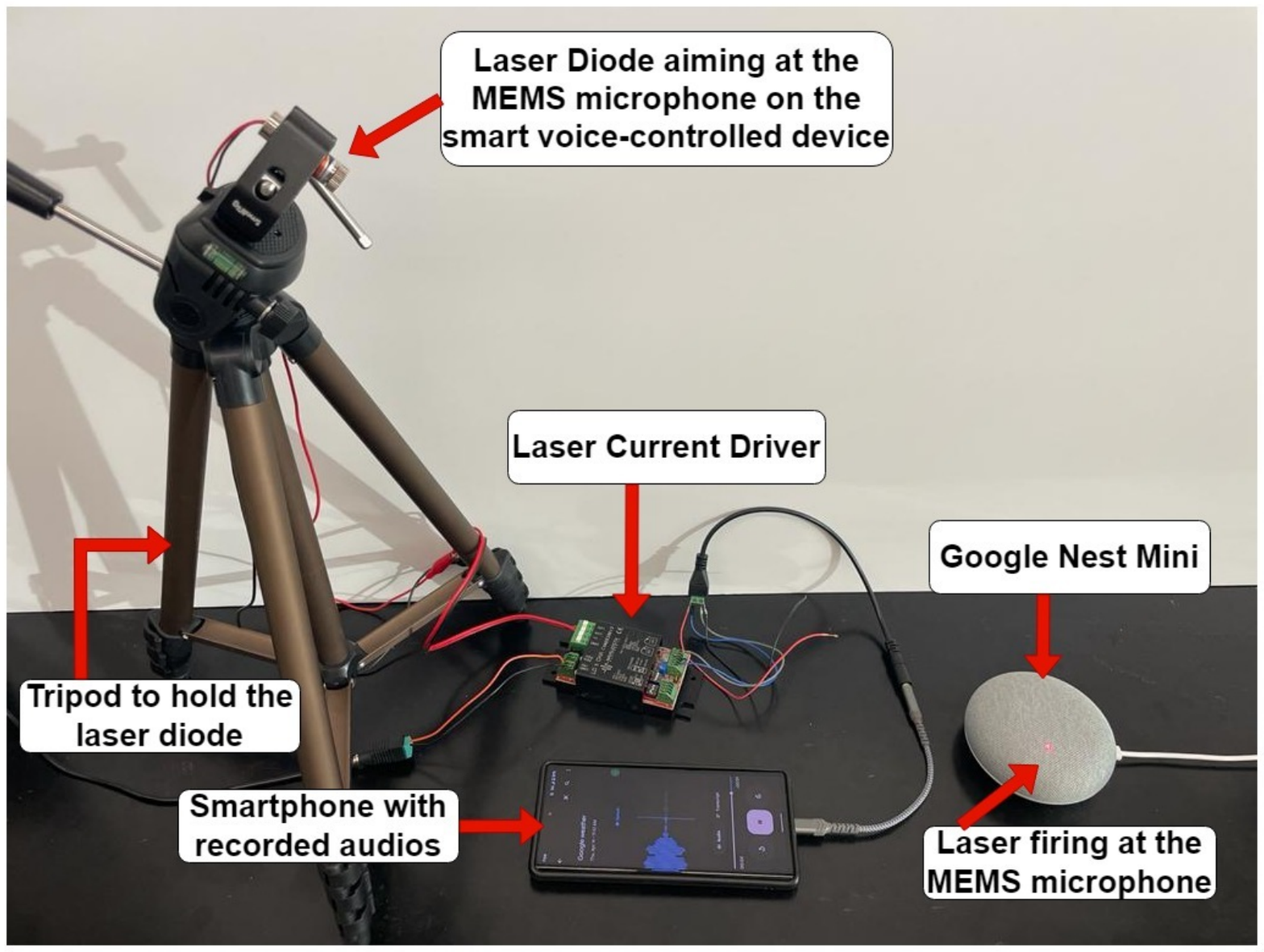}
     \caption{Experimental setup to replicate laser injection attack in VC Devices}
    \label{fig:exp_setup}
\end{figure}

To measure the effectiveness of the proposed ML-based countermeasure against laser injection attacks, we implemented the experimental setup proposed in \cite{sugawara2020light}. Shown in figure \ref{fig:exp_setup} is the implementation of the experimental setup for data collection and performance evaluation. This experimental setup is divided into two parts: (i)\textit{The Attacker Side} and (ii)\textit{The Victim Side}. \textit{The attacker side} consists of a laser pointer, laser current driver, 5V power supply, tripod for the laser, 3.5mm stereo audio cable, and connecting jumper wires. Although a laser diode is available in the range of \$5 - \$5000, we used a cheaper one for this experiment to demonstrate that even with a cheap laser diode, the attack is still possible. \textit{On the victim side}, the \textit{Google Nest Mini Kit} was used to replicate the Google Home Smart Assistant (refer to figure \ref{fig:exp_setup}).

This experimental setup is used to generate a dataset containing two types of audio recordings. The first set of audio recordings consists of MEMS microphone response to acoustic activities (e.g., 19 participants reading provided text in front of the Google Home Smart Assistant). The second set of audio recordings consists of MEMS microphone response to photo-acoustic activities (laser modulated--with audio recordings of 19 participants, firing at the MEMS microphone of Google Home Smart Assistant). A total of 19 students (10-male and 9-female) were enrolled for data collection. All participants were asked to read the following 5 sentences in the microphone, \textit{``Hey Google, Open the garage door"}, \textit{``Hey Google, Close the garage door"}, \textit{``Hey Google, Turn the light on"}, \textit{``Hey Google, Turn the light off"}, \textit{``Hey Google, What is the weather today?"}. Each audio sample was injected into the microphone through a laser, and the response of the microphone was recorded. This method produced a total data set of 95 acoustic- and 95 laser-induced audio recordings\textcolor{blue}{\footnote{https://www.kaggle.com/datasets/hashimali19/laser-injection-data}}.

To compare the effectiveness of the proposed higher-order statistical features of subband decomposition using DWT, three baseline features, CQCCs, LFCCs, and MFCCs (commonly used for replay attack detection) were considered. An SVM model with an RBF kernel is trained on these features. For this purpose, the Sklearn implementation of SVM \cite{scikit-learn} is used with default settings. Three experiments were performed to evaluate the robustness and reliability of the proposed framework.
% A linear SVM classifier with `l2' norm-based penalization and `hinge' loss is used to train on the collected dataset.  %For these experiments the dataset was segmented following three partitions: (i) random 70\% and 30\% partition for training and testing, (ii), and (iii). 
\subsubsection{Experiment 1: Speaker and Text Dependent Analysis} 
The goal of this experiment is to evaluate the performance of the proposed framework in a speaker-dependent and text-dependent setting. To achieve that goal, the complete data set was randomly partitioned into $70\%$ training and $30\%$ test data. This type of experiment is considered Speaker Dependent (SD) and Text Dependent (TD) as the training and test data may share the same speakers and text. The results of DWT, CQCC, LFCC, and MFCC features on SD and TD data partition are given in table \ref{table:table_comp}, column-1. It can be observed that both DWT and LFCC features were able to correctly classify $98\%$ of the acoustic-induced and laser-induced audio, whereas the classification accuracy for CQCC and MFCC features were $91\%$ and $96\%$ respectively.

 % the performance of the proposed framework in a speaker-independent and text-dependent setting
\subsubsection{Experiment 2: Speaker Independent (SI) and Text Dependent (TD) Analysis} 
The goal of this experiment is to investigate the impact of speaker-data leakage between train and test datasets on the detection performance of the proposed system. To achieve this goal, the complete data set, consisting of 19 speakers, was partitioned into two sets, namely Train and Test. The train set contains the first 14 speakers whereas the test set contains the remaining five speakers. Each speaker still has five utterances, the same in both the train and test sets. This type of experiment is considered speaker-independent (SI) as different sets of speakers are used for training and testing the model to avoid data leakage. The results of DWT, CQCC, LFCC, and MFCC features on SI data partition are given in table \ref{table:table_comp}, column-2. It can be observed that DWT, CQCC, and LFCC were able to correctly classify $100\%$ of the acoustic-induced and laser-induced audio, whereas the classification accuracy for MFCC features is $96\%$. 

% performance of the proposed framework in a speaker-independent and text-independent setting
\subsubsection{Experiment 3: Speaker Independent (SI) and Text Independent (TI) Analysis} 
The goal of this experiment is to investigate the impact of the text- and speaker-data leakage between train and test datasets on the detection performance of the proposed system. To achieve this goal, only the first three utterances of the first 14 speakers were used for model training and the remaining two utterances of the remaining 5 speakers were used for testing. The results of DWT, CQCC, LFCC, and MFCC features on SI + TI data partition are given in table \ref{table:table_comp}, column-3. It can be observed that DWT, CQCC, and LFCC were able to correctly classify $100\%$ of the acoustic-induced and laser-induced audio, whereas MFCC features were able to classify $95\%$ of the audio. Experiment 2 and 3 shows that the high accuracy of the proposed framework is not because of the data leakage between the train and test datasets.\textcolor{blue}{\footnote{https://github.com/hashim19/Laser\_Injection\_Attack\_Identification}}

\subsection{Frame-by-Frame Analysis} \label{Frame}
It is possible that a MEMS microphone can be attacked with audio containing laser-induced parts in it. To detect these types of attacks, a frame-by-frame analysis is performed using a sliding-window approach. Using a frame size $t_f$ of 1 sec and a hop length $t_h$ of 0.5 secs resulted in two types of frames: Non-Bordering frames (100\% laser-induced or acoustic-induced audio) and bordering frames (50\% laser-induced and 50\% acoustic-induced). DWT + SVM approach is able to detect 76\% of bordering and non-bordering frames whereas it is able to achieve an 80\% accuracy on only non-bordering frames.

\subsection{Robustness to Anti-Forensic Attack} \label{AFA}
\begin{figure}[h]
     \centering
     \includegraphics[width=1\linewidth]{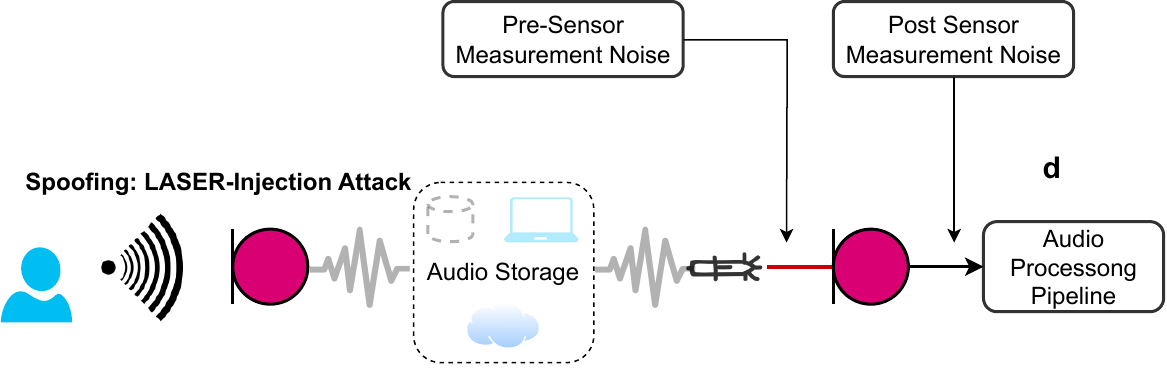}
     \caption{Two Scenarios for Anti-Forensics Attempts}
    \label{fig:antiForensic}
\end{figure}

The proposed framework leverages artifacts due to photo-acoustic excitation in the MEMS microphones for laser-injection attack detection. An attacker can craft an anti-forensic attack to bypass detection by either removing distinguishable artifacts. For example, an attacker can add color noise in low- and mid-frequency bands. There are two possible ways such an anti-forensic attack can be executed: (i) pre-sensor measurement noise addition, e.g., laser injection in a low-frequency background noise environment, and (ii) post-sensor measurement noise addition, e.g., a low-frequency noise addition into microphone output audio recording. Shown in figure \ref{fig:antiForensic} is the block diagram to execute the color noise addition attack. To determine the robustness of the proposed framework against such attacks we have started collecting data for pre-sensor measurement noise addition attack. Initial results indicate that the proposed method is robust to such anti-forensic attacks. This is mainly due to the fact the proposed system relies on characteristic artifacts due to photo-acoustic excitation to distinguish between laser- and acoustic-induced audio. These artifacts are independent of environmental acoustic activities and therefore robust to pre-sensor low-frequency noise addition attack scenarios. As far as the post-sensor measurement noise addition scenario is concerned, this attack vector requires sensor (MEMS microphone) access which means that the VAD is under attacker's control. This threat model is not considered here. %For the sake of completion, we strongly believe that the proposed system is expected to falsely label acoustic-induced audio as laser-induced audio when test audio is obtained by a carefully crafted noise profile added to the acoustic-induced audio. This attack is expected to increase false positive rates for the proposed system. 

% Standard Experimental Section Structure
% 1). What is the goal of the experiment
% 2). How we did it
% 3). The results of the experiment
% 4). Discussion/Analysis on the results

\section{Conclusion}
This paper investigated and developed a countermeasure for protecting Voice-Controlled devices against laser injection attacks. The proposed framework was able to correctly classify a data set of 190 acoustic- and laser-induced audios and performs at par with the baseline CQCC and LFCC features. However, the proposed framework is just a demonstration that laser-induced response in the microphone can be differentiated from acoustic-induced response. Other methods, as demonstrated above, may also work and may work even better than the proposed method. Therefore, a detailed analysis of existing countermeasures needs to be performed on a reasonably big amount of laser-injection data set. Moreover, the robustness of these countermeasures needs to be analyzed to colored noise addition in two scenarios, as described in section \ref{AFA}.

\bibliographystyle{IEEEtran}
\bibliography{references}

\end{document}